# A UNIFIED APPROACH TOWARDS DESCRIBING RAPIDITY AND TRANSVERSE MOMENTUM DISTRIBUTIONS IN THERMAL FREEZE-OUT MODEL

Saeed Uddin[1], Jan Shabir Ahmad[2], Waseem Bashir
and Riyaz Ahmad Bhat

Department of Physics, Jamia Millia Islamia, New Delhi - 110025

## ABSTRACT

We have attempted to describe the rapidity and transverse momentum spectra, simultaneously, of the hadrons produced in the Ultra-relativistic Nuclear Collisions. This we have tried to achieve in a single statistical thermal freeze-out model using single set of parameters. We assume the formation of a hadronic gas in thermo-chemical equilibrium at the freeze-out. The model incorporates a longitudinal as well as a transverse hydrodynamic flow. We have also found that the role of heavier hadronic resonance decay is important in explaining the particle spectra.

1   e-mail : saeed_jmi@yahoo.co.in
2   On leave of deputation from Amar Singh College, Srinagar, J & K

# INTRODUCTION

The yields of baryons and antibaryons are an important indicator of the multi-particle production phenomenon in the ultra-relativistic nucleus-nucleus collisions. The study of ultra-relativistic nuclear collisions allows us to learn how baryon numbers initially carried by the nucleons only, before the nuclear collision, are distributed in the final state [1] at the thermo-chemical freeze-out after the collision. The measurement of the *net* proton rapidity density distribution in such experiments can throw light on the collision scenario. At RHIC the midrapidity $p$ – pbar yield decreases gradually with increasing energy. Thus at RHIC energies the nuclear collisions start exhibiting some transparency [2-4] which clearly indicate that the antiproton to proton ratio shows a maximum at mid rapidity and gradually decreases towards larger rapidities, whereas the *net* proton rapidity density distribution shows a broad minimum, spanning about ± 1 unit around mid-rapidity region of dN/dy spectra. As the produced hadrons carry information about the collision dynamics and the entire space-time evolution of the system a precise measurement of the transverse momentum $p_T$ distributions and yields of identified hadrons along with the rapidity spectra is also essential for the understanding of the dynamics and properties of the created matter up to the thermo-chemical freeze-out.

Hydrodynamic models [5,6] that include radial flow successfully describe the measured $p_T$ distributions in Au+Au collisions at $\sqrt{s_{NN}}$=130 GeV [7-9]. The $p_T$ spectra of identified charged hadrons below $p_T$<2 GeV/*c* in central collisions have been well reproduced in some models by two simple parameters: transverse flow velocity $\beta_T$ and freeze-out temperature T [8], under the assumption of thermalization with longitudinal and transverse flow [5]. Some statistical thermal models have successfully described the particle abundances at low $p_T$ [10-12]. However a single model to describe rapidity and the $p_T$ distributions simultaneously is still lacking.

## THE MODEL

The hot and dense matter produced in relativistic heavy ion collisions may evolve through the following scenario :

Pre-equilibrium, thermal (or chemical) equilibrium of partons, possible formation of QGP or a QGP-hadron gas mixed state, a gas of hot interacting hadrons, a chemical freeze-out state when the produced hadrons no longer strongly interact with each other inelastically so that the particle number changing processes stop (except the resonance decay processes). After this the particles may continue to interact through elastic processes where a certain fraction of the thermal energy gets converted into a collective hydrodynamic flow thereby leading to a decrease in the local thermal temperatures. This would then lead to reshaping of the spectra of hadrons till a final stage where a ***hydrodynamic freeze-out*** would occur [13].

In our model it is assumed that a (near) simultaneous freeze-out of all hadronic species occurs.

In order to obtain the particle spectra in the overall rest frame of the hadronic fireball we first define the invariant cross-section for a given hadron in the local rest frame of the expanding hadronic fluid element. As the invariant cross-sections will have same shape in all Lorentz frames thus we can write :

$$E\frac{d^3n}{d^3p} = E'\frac{d^3n}{d^3p'} \qquad (1)$$

The primed quantities on the RHS refer to the invariant spectra of a given hadronic specie, in the local hadronic fluid element while the unprimed quantities on the LHS refer to the invariant spectra of the same hadronic specie but in the overall rest frame of the hadronic fireball, formed in the ultra-relativistic nuclear collisions. Further we can write :

$$E^{'} = m_{T}^{'} Cosh(y^{'}) \tag{2}$$

Where $m_{T}^{'} = \sqrt{p_{T}^{'2} + m^2}$ is the transverse mass and the $p_{T}^{'}$ and $y'$ are the transverse momentum and rapidity of the given hadron with mass $m$ in the rest frame of the local hadronic fluid element. We now use the quantum distribution function to write :

$$\frac{d^3 n}{d^3 p^{'}} \sim \frac{1}{\exp[(E^{'} - \mu)/T] \pm 1}$$

where $\mu$ is the chemical potential of the given hadronic specie. We now apply Lorentz transformations to write the primed quantities on the RHS in Eqn. (1) in terms of the unprimed quantities as follows :

$$p_{T}^{'} = \gamma_{T}(p_{T} - \beta_{T} E) \quad \text{and} \quad y^{'} = y - y_{0}$$

where $y_0$ is the rapidity of the local hadronic fluid element in the overall rest frame of the hadronic fireball.

We have assumed that $y_0 \propto z$, where z represents the z-coordinate of the fluid element in the rest frame of the hadronic fireball. Thus writing $y_0 = cz$, we obtain the following expression for the longitudinal velocity component of the hadronic fluid element :

$$\beta_z(z) = 1 - \frac{2}{\exp(2cz)+1} \qquad (3)$$

In recent works [10, 14] it has been clearly shown that there is a strong evidence of increasing baryon chemical potential along the collision axis. It is therefore possible to reproduce the variation of the antibaryon to baryon ratio with rapidity in accordance with the RHIC data only when we incorporate this effect [13]. Hence in the present model we incorporate this effect by writing :

$$\mu = a + b y_0^2 \qquad (4)$$

Where $y_0$ is the rapidity of the hadronic fluid element along the beam axis. Further as in many earlier works it is assumed here also that the transverse (or radial) velocity component of the hadronic fluid varies with the transverse coordinate $r$ as [13, and the references therein] :

$$\beta_T(r) = \beta_T^S (r/R)^n \tag{5}$$

where $R$ is the transverse radius of the fireball and $n$ is an index to fix the profile of $\beta_T(r)$ in the radial direction. In order to reproduce the correct observed rapidity distributions of hadrons it is necessary in the present model to assume that the transverse radius of the fireball decreases with the z coordinate of the system as :

$$R = r_0 \exp(-|z|^2/\sigma^2) \tag{6}$$

Consequently at the freeze-out we don't have a spherically symmetric system but rather a system which with varying transverse radius follows a Gaussian profile along the rapidity (or beam) axis. The other parameter viz. the fluid's surface transverse expansion velocity $\beta_T^S$ is fixed in the model by using the parameterization :

$$\beta_T^S = \beta_T^0[1 - \beta_z^2(z)] \qquad (7)$$

The above relation (or restriction) is also required to ensure that the net particle velocity β must satisfy :

$$\beta = \sqrt{\beta_T^2 + \beta_z^2} < 1 \qquad (8)$$

We finally perform an integral over the physical volume of the system to obtain the net hadronic yield (proton and antiprotons here).

## RESULTS

In figure 1 and 2 we have shown the rapidity spectra of protons and antiprotons. On the same we have also shown the BRAHMS experimental data points obtained from the most central Au + Au collisions at √$s_{NN}$=200 GeV at RHIC. The parameter values used to obtain these curves are same. We have used T= 150 MeV, a = 20.5 MeV, b = 10.4 MeV, c = 1 fm$^{-1}$, σ = 4.2 fm$^2$, n = 2 and $\beta_T^0 = 0.72$

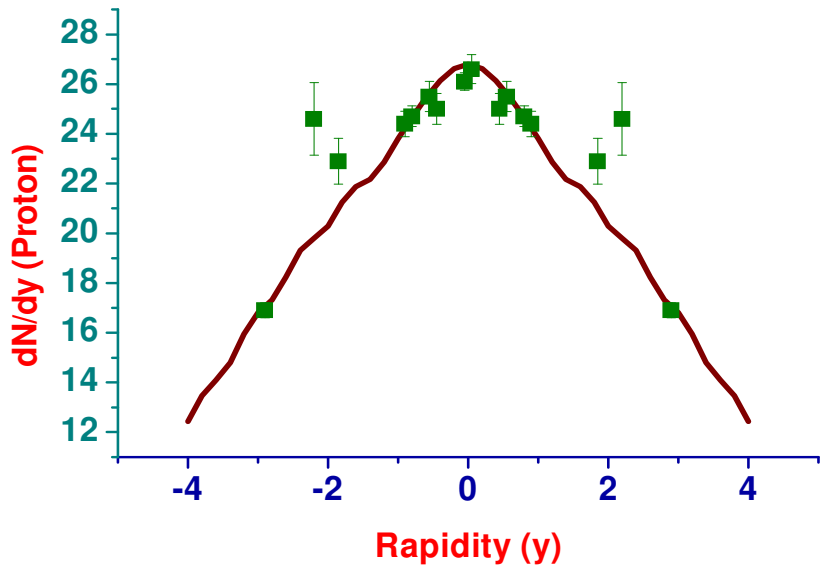

FIGURE 1 : Rapidity distribution of protons. The BRAHMS data points are shown by green boxes.

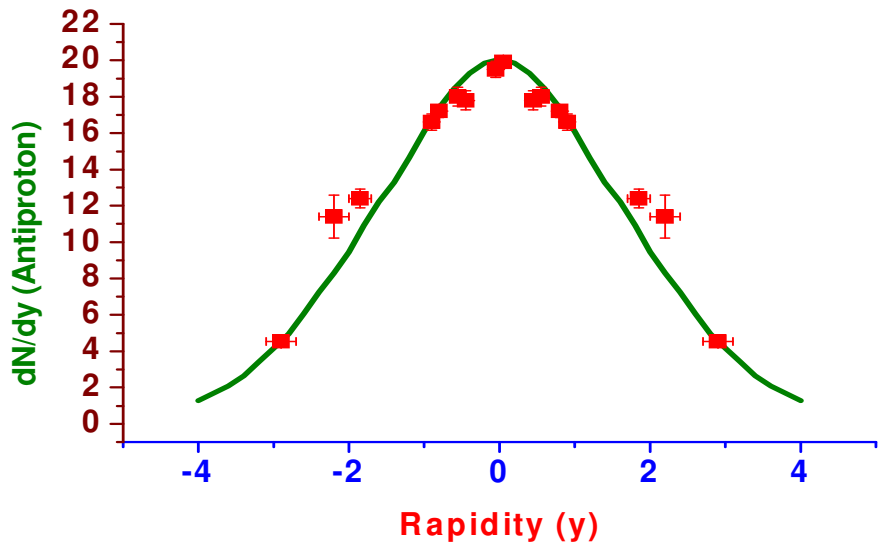

FIGURE 2 : Rapidity distribution of antiprotons. The BRAHMS data points are shown by red boxes.

In Figures 3 and 4 we have shown our theoretical results for the *net* proton density distribution and the pbar/p ratio. We see that the theoretical curves fit the experimental data points very well in all the above cases. In Figure 4 the effect of increasing chemical potential along the rapidity axis is clearly exhibited by the falling pbar/p ratio with the rapidity.

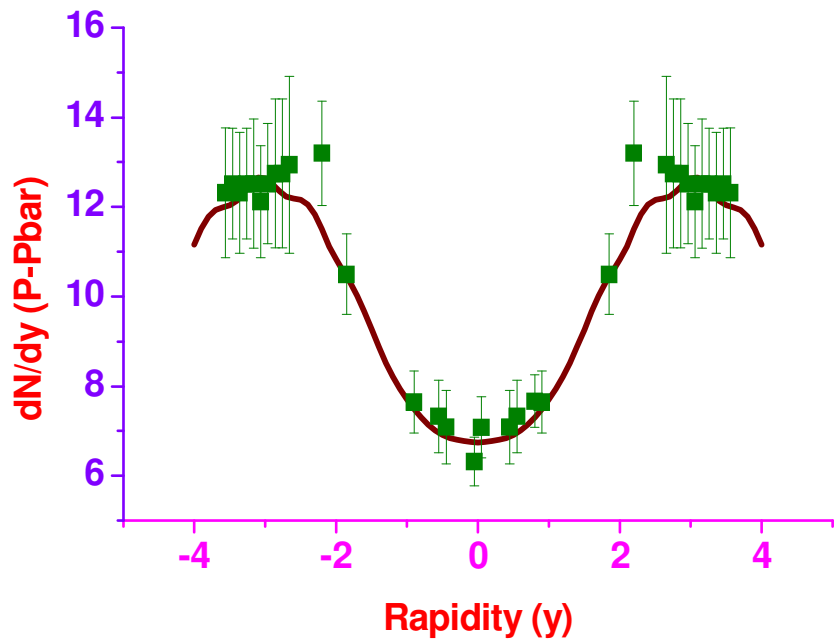

FIGURE 3 : Rapidity distribution of net protons. The data points from BRAHMS are shown by green boxes.

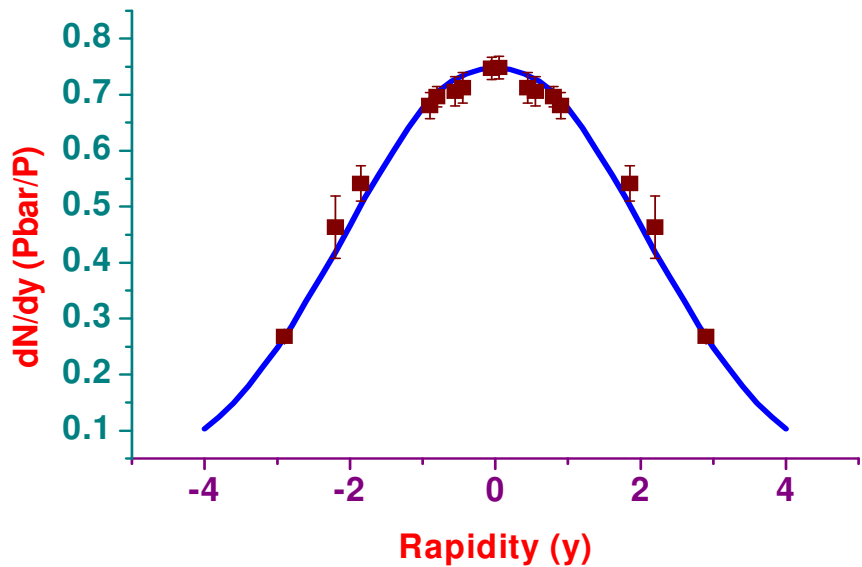

FIGURE 4 : Rapidity distribution of pbar/p. The boxes show the experimental data points of BRAHMS

Further in Figures 5 and 6 we have shown the transverse momentum spectra of the protons and antiprotons, respectively. We again find that the theoretical curves fit the experimental data points quite well. The experimental data points used here are from the PHENIX. The BRAHMS rapidity spectra data as well as the PHENIX transverse momentum spectra data sets are for the 0 – 5 % most central collisions of Au + Au at $\sqrt{s_{NN}}$=200 GeV [15]. It may be noted that the theoretical curves in all the six cases have been obtained for the same set of parameters described above. In all these calculations we have also incorporated the contribution of the resonance decays. We have found that it is necessary to incorporate the contribution of the decaying hadrons as without this we can not obtain the correct rapidity and the transverse momentum distributions obtained in the RHIC experiments. This is in contrast to earlier studies [14, and the references therein] where the transverse flow effect was completely ignored in the model to describe the rapidity spectra of various hadrons.

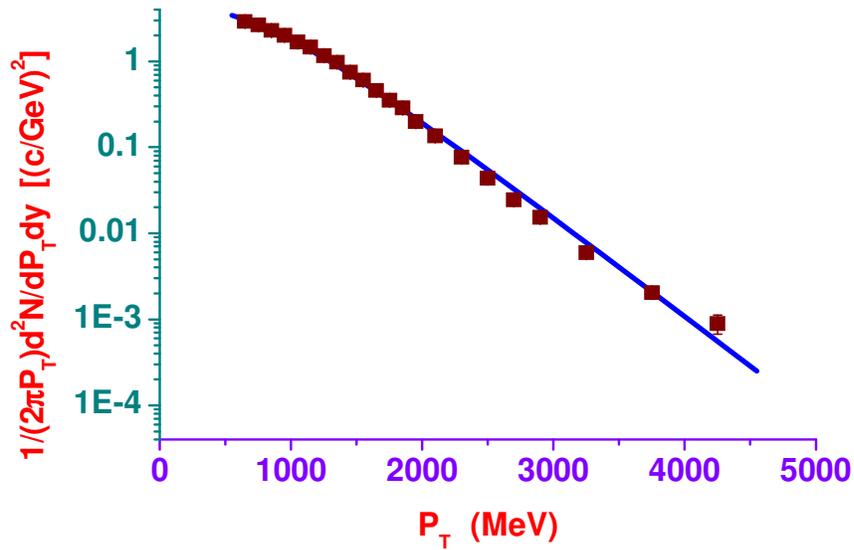

FIGURE 5 : The transverse momentum ($P_T$) distribution of protons. The boxes show the experimental data points from PHENIX

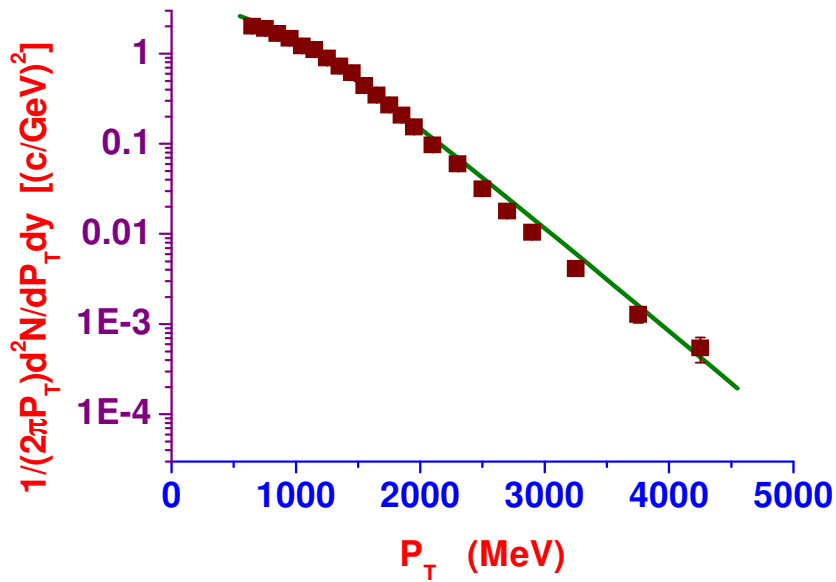

FIGURE 6 : The transverse momentum ($P_T$) distribution of antiprotons. The boxes show the experimental data points from PHENIX

In the present model we find that the resonance decay plays a crucial role in (re)shaping the rapidity spectra. To highlight this effect we have plotted the rapidity spectra of protons and antiprotons with and without the resonance decay contributions in Figures 7 and 8, respectively. In both cases we find that there is a significant dip in the dN/dy values at midrapidity. This indicates that the contribution of the decay particles is very significant at smaller rapidity values (y,1), while at larger rapidities (y>1.5) the decay contributions become negligible, which is evident by the merging of the two curves (i.e. with and without resonance decay contributions).

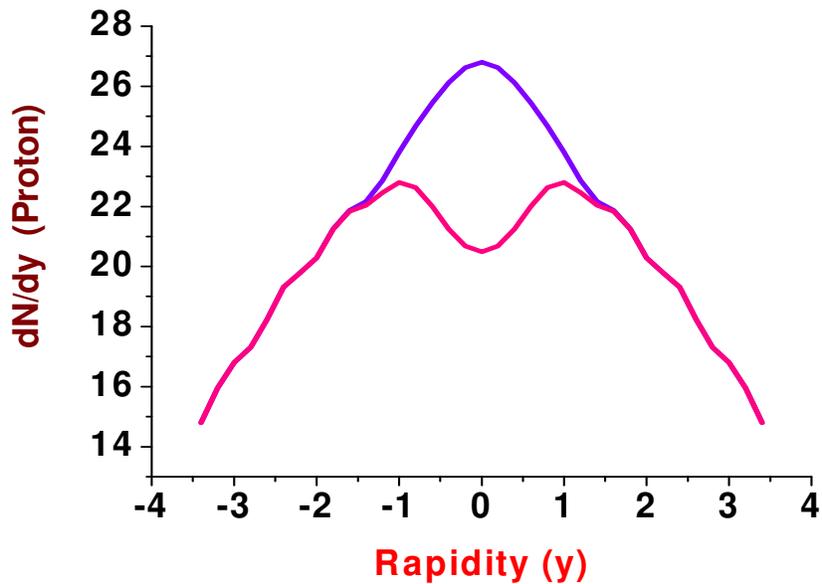

FIGURE 7 : Theoretical rapidity distribution of Protons. The violet curve is for the case where resonance decay contributions are taken into account, while the pink curve is for the case where decay contributions are ignored.

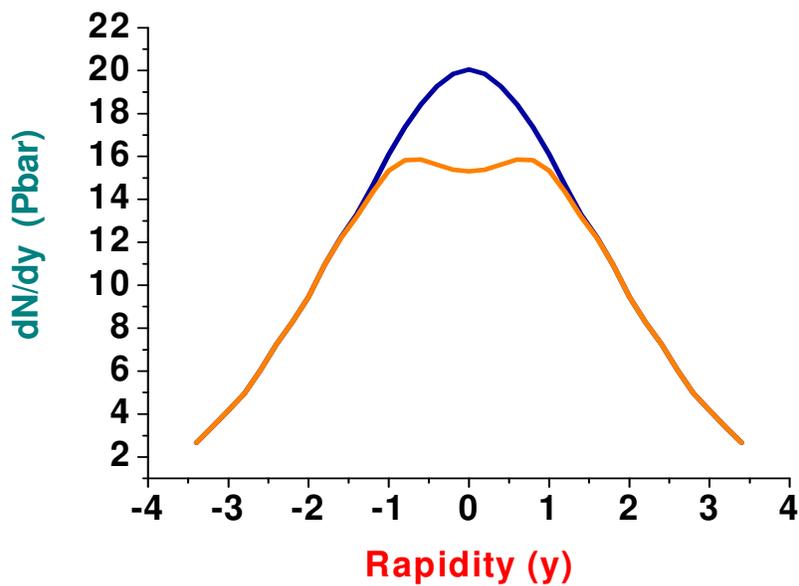

FIGURE 7 : Theoretical rapidity distribution of Antiprotons. The blue curve is for the case where resonance decay contributions are taken into account, while the orange curve is for the case where decay contributions are ignored.

However, interestingly, a similar analysis to assess the significance of the resonance decay contributions in the transverse momentum distributions of protons and antiprotons reveals that the $P_T$ spectra of protons and antiprotons are almost unchanged if we do not included the decay contributions, i.e. if we consider only the directly produced protons and antiprotons. This is clearly highlighted by the two merging curves in the

Figure 8 for protons and Figure 9 for antiprotons. Note that the two curves for the cases of with and without resonance decay contributions have been normalized at $p_T$ = 550 MeV to facilitate a proper comparison of their shapes in both the Figures.

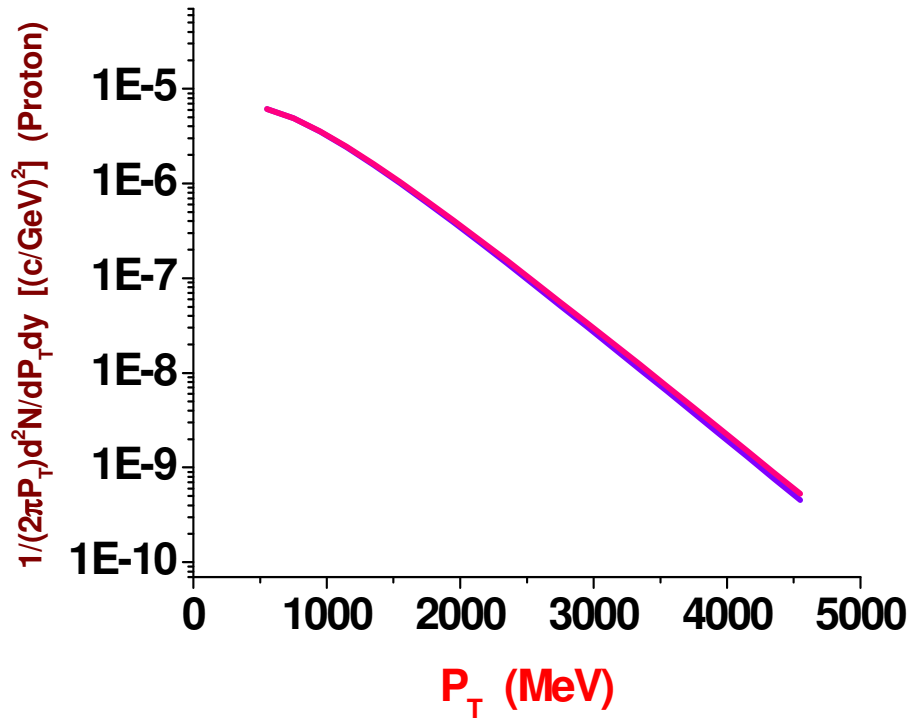

FIGURE 8 : The transverse momentum ($P_T$) distribution of protons. The two (nearly overlapping) curves are for the two cases, with resonance contributions taken into account (violet curve) and without resonance contribution (pink curve).

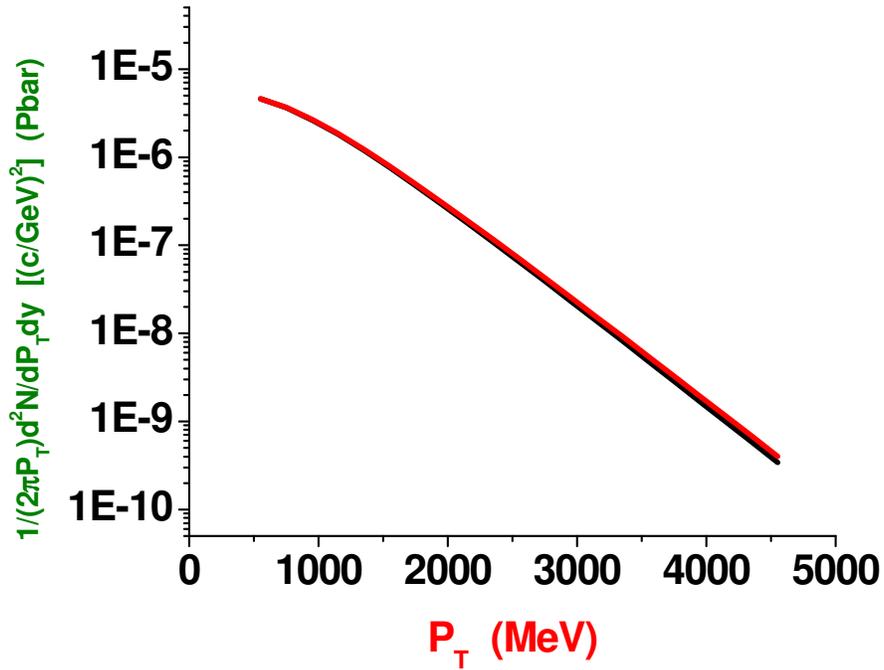

FIGURE 9 : The transverse momentum ($P_T$) distribution of antiprotons. The two (nearly overlapping) curves are for the two cases, with resonance contributions taken into account (black curve) and without resonance contribution (red curve).

Further we have also attempted to study the effect of the transverse surface expansion velocity parameter namely $\beta_T^0$. In Figure 10 we have shown the $p_T$ spectrum of protons (with resonance decay contribution included) for three different values of $\beta_T^0$ = 0.62, 0.72 and 0.82. It is seen that as the value of $\beta_T^0$ increases the spectrum becomes broader, thereby showing a larger apparent temperature, T$^{app}$ (or the inverse slope parameter).

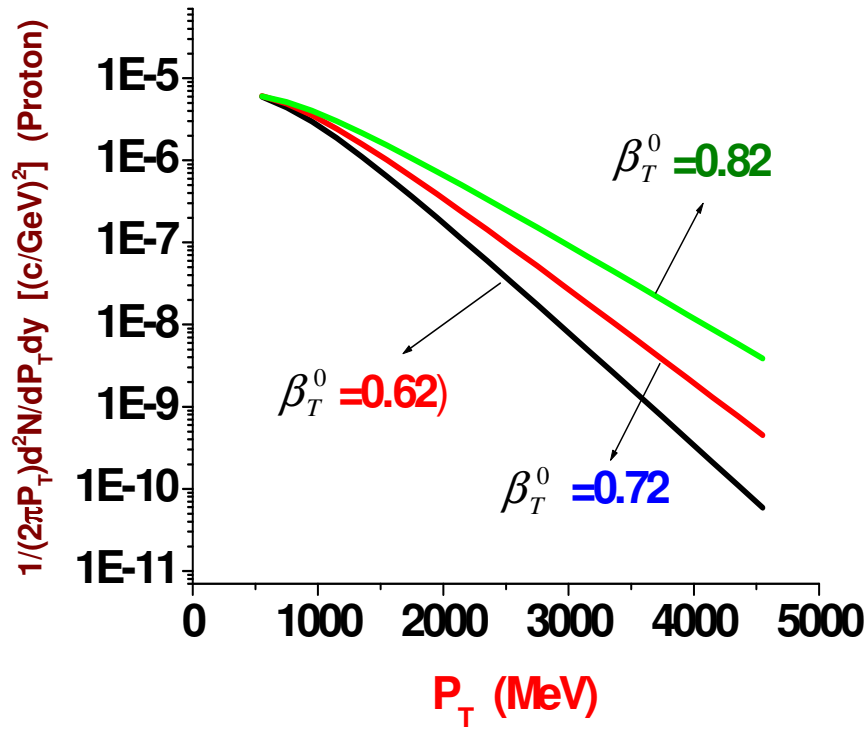

FIGURE 10 : The transverse momentum (P$_T$) distribution of protons corresponding to three different values of $\beta_T^0$ show larger apparent temperature due to the broading of the spectral shapes at higher values of tansverse momentum.

We have further highlighted the effect of the flow velocity in reshaping the particles spectra as the collective flow gives the particle spectra a higher apparent temperature, T$^{app}$, than the

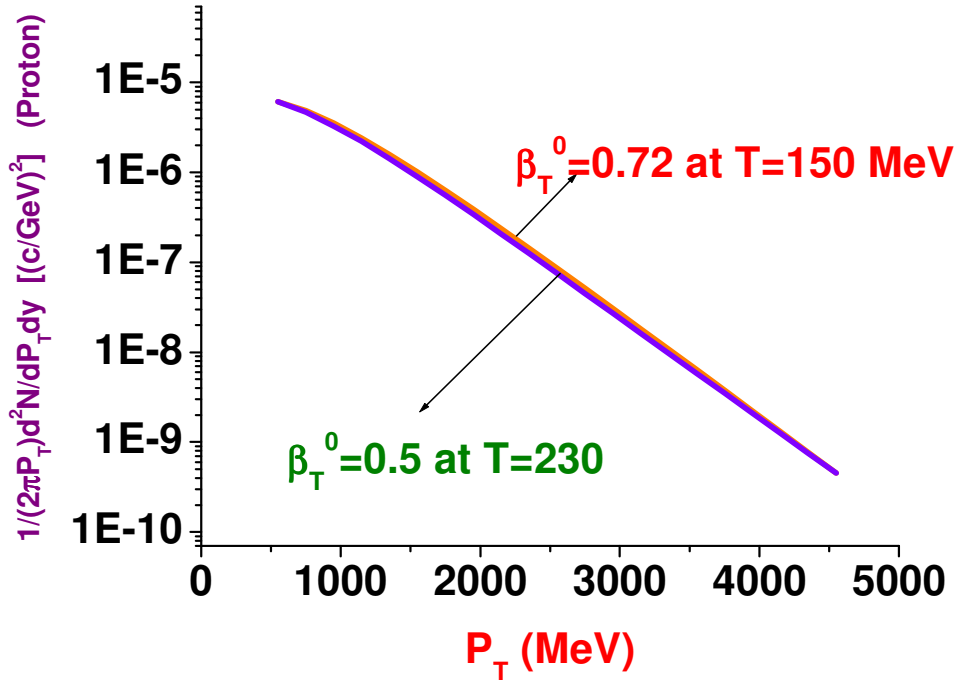

FIGURE 11 : The transverse momentum ($P_T$) distribution of protons. The two (nearly overlapping) curves are for the two cases, with $\beta_T^0 = 0.72$ and T= 150 MeV (orange curve) and with $\beta_T^0 = 0.5$ and T=230 MeV (violet curve).

actual thermal temperature, T. In Figure 11 we have plotted two curves. In one case the thermal temperature of proton is raised to 230 MeV after decreasing the value of $\beta_T^0$ to 0.5, while the other case is for T= 150 MeV and $\beta_T^0$ = 0.72. We find that the two curves have the same shape. Thus an apparently large value of temperature, $T^{app}$, can be obtained at a much lower thermal temperature if we superimpose a strong collective flow over the thermal spectra. However, our choice of the lower thermal temperature is guided by some other particle ratios such as Ξ/p, etc. We are presently calculating many such particle ratios as well as the spectra of pions and Kaons.

## SUMMARY AND CONCLUSIONS

We have suitably modified the existing thermal model to describe the rapidity as well as transverse momentum spectra of hadrons in a single thermal freeze-out model. This is achieved by incorporating a longitudinal as well as a transverse flow. The system size is assumed to decrease in the transverse directions following a Gaussian profile in the z-coordinate. The model provides a very good description of the rapidity and transverse momentum spectra of protons and antiprotons obtained from the 0 – 5% most central Au + Au collisions at $\sqrt{s_{NN}}$=200 at RHIC.